\documentclass[twocolumn,showpacs,superscriptaddress,preprintnumbers,amsmath,amssymb]{revtex4}
\usepackage{graphicx}
\usepackage{dcolumn}
\usepackage{bm}
\begin{document}


\title{Signatures of the $d^*(2380)$ hexaquark in d($\gamma$,$p\vec{n}$)}
\author{M. Bashkanov}
 \affiliation{Department of Physics, University of York, Heslington, York, Y010 5DD, UK}
\author{D.P. Watts}%
\email{daniel.watts@york.ac.uk}
\affiliation{Department of Physics, University of York, Heslington, York, Y010 5DD, UK}

\author{S.J.D.~Kay}
\affiliation{University of Regina, Regina, SK S4S0A2 Canada}
\author{S.~Abt}
\affiliation{Department of Physics, University of Basel, Ch-4056 Basel, Switzerland}
\author{P.~Achenbach}
\affiliation{Institut f\"ur Kernphysik, University of Mainz, D-55099 Mainz, Germany}
\author{P.~Adlarson}
\affiliation{Institut f\"ur Kernphysik, University of Mainz, D-55099 Mainz, Germany}
\author{F.~Afzal}
\affiliation{Helmholtz-Institut f\"ur Strahlen- und Kernphysik, University Bonn, D-53115 Bonn, Germany}
\author{Z.~Ahmed}
\affiliation{University of Regina, Regina, SK S4S0A2 Canada}
\author{C.S.~Akondi}
\affiliation{Kent State University, Kent, Ohio 44242, USA}
\author{J.R.M.~Annand}
\affiliation{SUPA School of Physics and Astronomy, University of Glasgow, Glasgow, G12 8QQ, UK}
\author{H.J.~Arends}
\affiliation{Institut f\"ur Kernphysik, University of Mainz, D-55099 Mainz, Germany}
\author{R.~Beck}
\affiliation{Helmholtz-Institut f\"ur Strahlen- und Kernphysik, University Bonn, D-53115 Bonn, Germany}
\author{M.~Biroth}
\affiliation{Institut f\"ur Kernphysik, University of Mainz, D-55099 Mainz, Germany}
\author{N.~Borisov}
\affiliation{Joint Institute for Nuclear Research, 141980 Dubna, Russia}  
\author{A.~Braghieri}
\affiliation{INFN Sezione di Pavia, I-27100 Pavia, Pavia, Italy}
\author{W.J.~Briscoe}
\affiliation{Center for Nuclear Studies, The George Washington University, Washington, DC 20052, USA}
%
\author{F.~Cividini}
\affiliation{Institut f\"ur Kernphysik, University of Mainz, D-55099 Mainz, Germany}
\author{C.~Collicott}
\affiliation{Department of Astronomy and Physics, Saint Mary's University, E4L1E6 Halifax, Canada}
\author{S.~Costanza}
\affiliation{Dipartimento di Fisica, Universit\`a di Pavia, I-27100 Pavia, Italy}
\affiliation{INFN Sezione di Pavia, I-27100 Pavia, Pavia, Italy}
\author{A.~Denig}
\affiliation{Institut f\"ur Kernphysik, University of Mainz, D-55099 Mainz, Germany}
\author{E.J.~Downie}
\affiliation{Center for Nuclear Studies, The George Washington University, Washington, DC 20052, USA}
\author{P.~Drexler}
\affiliation{Institut f\"ur Kernphysik, University of Mainz, D-55099 Mainz, Germany}
\affiliation{II. Physikalisches Institut, University of Giessen, D-35392 Giessen, Germany}
%
%
\author{S.~Gardner}
\affiliation{SUPA School of Physics and Astronomy, University of Glasgow, Glasgow, G12 8QQ, UK}
\author{D.~Ghosal}
\affiliation{Department of Physics, University of Basel, Ch-4056 Basel, Switzerland}  
\author{D.I.~Glazier}
\affiliation{SUPA School of Physics and Astronomy, University of Glasgow, Glasgow, G12 8QQ, UK}
\author{I.~Gorodnov}
\affiliation{Joint Institute for Nuclear Research, 141980 Dubna, Russia}
\author{W.~Gradl}
\affiliation{Institut f\"ur Kernphysik, University of Mainz, D-55099 Mainz, Germany}
\author{M.~G\"unther}
\affiliation{Department of Physics, University of Basel, Ch-4056 Basel, Switzerland}  
\author{D.~Gurevich}
\affiliation{Institute for Nuclear Research, RU-125047 Moscow, Russia}
\author{L. Heijkenskj{\"o}ld}
\affiliation{Institut f\"ur Kernphysik, University of Mainz, D-55099 Mainz, Germany}
\author{D.~Hornidge}
\affiliation{Mount Allison University, Sackville, New Brunswick E4L1E6, Canada}
\author{G.M.~Huber}
\affiliation{University of Regina, Regina, SK S4S0A2 Canada}
\author{A.~K{\"a}ser}
\affiliation{Department of Physics, University of Basel, Ch-4056 Basel, Switzerland}   
\author{V.L.~Kashevarov}
\affiliation{Institut f\"ur Kernphysik, University of Mainz, D-55099 Mainz, Germany}
\affiliation{Joint Institute for Nuclear Research, 141980 Dubna, Russia}

\author{M.~Korolija}
\affiliation{Rudjer Boskovic Institute, HR-10000 Zagreb, Croatia}
\author{B.~Krusche}
\affiliation{Department of Physics, University of Basel, Ch-4056 Basel, Switzerland}
\author{A.~Lazarev}
\affiliation{Joint Institute for Nuclear Research, 141980 Dubna, Russia}  
\author{K.~Livingston}
\affiliation{SUPA School of Physics and Astronomy, University of Glasgow, Glasgow, G12 8QQ, UK}
\author{S.~Lutterer}
\affiliation{Department of Physics, University of Basel, Ch-4056 Basel, Switzerland}
\author{I.J.D.~MacGregor}
\affiliation{SUPA School of Physics and Astronomy, University of Glasgow, Glasgow, G12 8QQ, UK}
\author{D.M.~Manley}
\affiliation{Kent State University, Kent, Ohio 44242, USA}
\author{P.P.~Martel}
\affiliation{Institut f\"ur Kernphysik, University of Mainz, D-55099 Mainz, Germany}
\affiliation{Mount Allison University, Sackville, New Brunswick E4L1E6, Canada}
\author{R.~Miskimen}
\affiliation{University of Massachusetts, Amherst, Massachusetts 01003, USA}
\author{E.~Mornacchi}
\affiliation{Institut f\"ur Kernphysik, University of Mainz, D-55099 Mainz, Germany}
%
\author{C. Mullen}
\affiliation{SUPA School of Physics and Astronomy, University of Glasgow, Glasgow, G12 8QQ, UK}
\author{A.~Neganov}
\affiliation{Joint Institute for Nuclear Research, 141980 Dubna, Russia}  
\author{A.~Neiser}
\affiliation{Institut f\"ur Kernphysik, University of Mainz, D-55099 Mainz, Germany}
%
\author{M.~Ostrick}
\affiliation{Institut f\"ur Kernphysik, University of Mainz, D-55099 Mainz, Germany}
\author{P.B.~Otte}
\affiliation{Institut f\"ur Kernphysik, University of Mainz, D-55099 Mainz, Germany}
%
\author{D.~Paudyal}
\affiliation{University of Regina, Regina, SK S4S0A2 Canada}
\author{P.~Pedroni}
\affiliation{INFN Sezione di Pavia, I-27100 Pavia, Pavia, Italy}

\author{A. Powell}
\affiliation{SUPA School of Physics and Astronomy, University of Glasgow, Glasgow, G12 8QQ, UK}

\author{S.N.~Prakhov}

\affiliation{University of California Los Angeles, Los Angeles, California 90095-1547, USA}
%
\author{G.~Ron}
\affiliation{Racah Institute of Physics, Hebrew University of Jerusalem, Jerusalem 91904, Israel}
%
\author{A.~Sarty}
\affiliation{Department of Astronomy and Physics, Saint Mary's University, E4L1E6 Halifax, Canada}
\author{C.~Sfienti}
\affiliation{Institut f\"ur Kernphysik, University of Mainz, D-55099 Mainz, Germany}
\author{V.~Sokhoyan}
\affiliation{Institut f\"ur Kernphysik, University of Mainz, D-55099 Mainz, Germany}
\author{K.~Spieker}
\affiliation{Helmholtz-Institut f\"ur Strahlen- und Kernphysik, University Bonn, D-53115 Bonn, Germany}
\author{O.~Steffen}
\affiliation{Institut f\"ur Kernphysik, University of Mainz, D-55099 Mainz, Germany}
\author{I.I.~Strakovsky}
\affiliation{Center for Nuclear Studies, The George Washington University, Washington, DC 20052, USA}
\author{T.~Strub}
\affiliation{Department of Physics, University of Basel, Ch-4056 Basel, Switzerland}
\author{I.~Supek}
\affiliation{Rudjer Boskovic Institute, HR-10000 Zagreb, Croatia}
\author{A.~Thiel}
\affiliation{Helmholtz-Institut f\"ur Strahlen- und Kernphysik, University Bonn, D-53115 Bonn, Germany}
\author{M.~Thiel}
\affiliation{Institut f\"ur Kernphysik, University of Mainz, D-55099 Mainz, Germany}
\author{A.~Thomas}
\affiliation{Institut f\"ur Kernphysik, University of Mainz, D-55099 Mainz, Germany}
%
\author{Yu.A.~Usov}
\affiliation{Joint Institute for Nuclear Research, 141980 Dubna, Russia}  
\author{S.~Wagner}
\affiliation{Institut f\"ur Kernphysik, University of Mainz, D-55099 Mainz, Germany}
\author{N.K.~Walford}
\affiliation{Department of Physics, University of Basel, Ch-4056 Basel, Switzerland}
\author{D.~Werthm\"uller}
\affiliation{Department of Physics, University of York, Heslington, York, Y010 5DD, UK}
\author{J.~Wettig}
\affiliation{Institut f\"ur Kernphysik, University of Mainz, D-55099 Mainz, Germany}
\author{M.~Wolfes}
\affiliation{Institut f\"ur Kernphysik, University of Mainz, D-55099 Mainz, Germany}
\author{N.~Zachariou}
\affiliation{Department of Physics, University of York, Heslington, York, Y010 5DD, UK}
\author{L.A.~Zana}
\affiliation{Thomas Jefferson National Accelerator Facility, Newport News, VA 23606, USA}

\collaboration{A2 Collaboration at MAMI}

\date{\today}

\begin{abstract}
We report a measurement of the spin polarisation of the recoiling neutron in deuterium photodisintegration, utilising a new large acceptance polarimeter within the Crystal Ball at MAMI. The measured photon energy range of 300~--~700~MeV provides the first measurement of recoil neutron polarisation at photon energies where the quark substructure of the deuteron plays a role, thereby providing important new constraints on photodisintegration mechanisms. A very high neutron polarisation in a narrow structure centred around $E_{\gamma}\sim$~570~MeV is observed, which is inconsistent with current theoretical predictions employing nucleon resonance degrees of freedom. A Legendre polynomial decomposition suggests this behaviour could be related to the excitation of the $d^*(2380)$ hexaquark.
\end{abstract}

\maketitle


\section{\label{sec:Intro} Introduction}
The photodisintegration of the deuteron is one of the simplest reactions in nuclear physics, in which a well understood and clean electromagnetic probe leads to the breakup of a few-body nucleonic system. However, despite experimental measurements of deuteron photodisintegration spanning almost a century~\cite{Chadwick}, many key experimental observables remain unmeasured. This is particularly evidenced at distance scales (photon energies) where the quark substructure of the deuteron can be excited. This limits a detailed assessment of the reaction mechanism,  including the contributions of nucleon resonances and meson exchange currents as well as potential roles for more exotic QCD possibilities, such as the $d^*(2380)$ hexaquark recently evidenced in a range of nucleon-nucleon scattering reactions~\cite{mb,MB,MBC,TS1,TS2,MBA,MBE1,MBE2}. The $d^*(2380)$ has inferred quantum numbers $I(J^P)=0(3^+)$ and a mass $\sim~2380$~MeV, which in photoreactions would correspond to a pole at $E_{\gamma}\sim$~570~MeV. Constraints on the existence, properties end electromagnetic coupling of the $d^*(2380)$ would have important ramifications for the emerging field of non-standard multiquark states and our understanding of the dynamics of condensed matter systems such as neutron stars~\cite{nstars}.

Although cross sections for deuterium photodisintegration have been determined~\cite{DAPHNE}, polarisation observables provide different sensitivities to the underlying reaction processes and are indispensable in constraining the basic photoreaction amplitudes. Of all the single-polarisation variables, the ejected nucleon polarisation ($P_{y}$) is probably the most challenging experimentally, requiring the characterisation of a sufficient statistical quantity of events where the ejectile nucleon subsequently undergoes a (spin-dependent) nuclear scattering reaction in an analysing medium. Nucleon polarisation measurements of sufficient quality have therefore only recently become feasible with the availability of sufficiently intense photon beams. Efforts to date have focused on recoil proton polarisation ($P_{y}^p$)~\cite{JLabP,TOK1}, exploiting proton polarimeters in the focal planes of (small acceptance) magnetic spectrometers. The data have good statistical accuracy but with a discrete and sparse coverage of incident photon energy and breakup kinematics~\cite{JLabP,TOK1}, with most data restricted to a proton polar angle of $\Theta_p^{CM}\sim~90^{\circ}$ in the  photon-deuteron centre-of-mass (CM) frame. However, these available $P_{y}^p$ data do exhibit a distinct behaviour, reaching $\sim~-1$ (i.e. around -100\% polarisation), in a narrow structure centred on $E_{\gamma}\sim$~550~MeV. Due to the inability to describe this behaviour with theoretical calculations including only the established nucleon resonances, it was speculated~\cite{TOK1,TOK2} that it would be consistent with a then unknown 6-quark resonance, with inferred properties having a striking similarity to the $d^*(2380)$ hexaquark discovered later in $NN$ scattering. 
 
 Clearly, measurement of the ejected neutron polarisation ($P_{y}^n$) would be important to establish a role for the $d^*(2380)$ in photodisintegration. In  $d^*(2380)\rightarrow pn$ decays, the spins of the proton and neutron would be expected to be aligned \footnote{The spin 3 nature of the $d^*(2380)$ requires high partial waves in the decay to a proton-neutron final state ~\cite{MBE1,MBE2}. In 90\% of cases this is via the $^3D_3$ partial wave (angular momentum $L=2$, nucleon spins and $L$ all aligned) or in 10\% of cases via the $^3G_3$ partial wave (angular momentum $L=4$, nucleon spins aligned, spin and $L$ anti-aligned).}. Therefore, if the $P_{y}^p$ anomaly originates from a $d^*(2380)$ contribution, the neutron polarisation should mimic this anomalous behaviour.   
Measurements of $P_{y}^n$ are even more challenging experimentally than $P_{y}^p$, due to the inability to track the uncharged neutron into the scattering medium, and have only been obtained below $E_{\gamma}\sim$~30~MeV~\cite{previous_py1,previous_py2}. The experimental difficulties even led to attempts to extract $P_{y}^n$ from studies of the inverse reaction $\vec{n}+p \rightarrow d + \gamma$, using detailed balance~\cite{inversePy1,inversePy2}.

This new work provides the first measurement of $P_{y}^n$ in deuterium photodisintegration for $E_{\gamma}$ sensitive to the quark substructure of the deuteron, covering $E_{\gamma}=300-700$~MeV and neutron breakup angles in the photon-deuteron CM frame of $\Theta_{n}^{CM}=60-120^{\circ}$. 
  




\section{Experimental Details}
The measurement employed a new large acceptance neutron polarimeter~\cite{proposal} within the Crystal Ball detector at the A2@MAMI~\cite{MAMI} facility during a 300 hour beamtime. An 1557~MeV longitudinally polarised electron beam impinged on either a thin amorphous (cobalt-iron alloy) or crystalline (diamond) radiator, producing circularly (alloy) or linearly (diamond) polarised bremsstrahlung photons. As photon beam polarisation is not used to extract $P_{y}^{n}$, equal flux from the two linear/circular polarisation settings were combined to increase the unpolarised yield. The photons were energy-tagged ($\Delta E\sim~2$~MeV) by the Glasgow-Mainz Tagger~\cite{Tagg} and impinged on a 10 cm long liquid deuterium target cell.  Reaction products were detected by the Crystal Ball (CB)~\cite{CB}, a highly segmented NaI(Tl) photon calorimeter covering nearly 96\% of $4\pi$ steradians. For this experiment, a new bespoke 24 element, 7~cm diameter and  30~cm long plastic scintillator barrel (PID-POL)~\cite{PID} surrounded the target, with a smaller diameter than the earlier PID detector~\cite{PID}, but provided similar particle identification capabilities. A 2.6~cm thick cylinder of analysing material (graphite) for nucleon polarimetry was placed around PID-POL, covering polar angles $12^{\circ} < \theta < 150^{\circ}$ and occupying the space between PID-POL and the Multi Wire Proportional Chamber (MWPC)~\cite{MWPC}. The MWPC provided charged particle tracking for particles passing out of the graphite into the CB. At forward angles, an additional 2.6~cm thick graphite disc covered the range $2< \theta < 12^{\circ}$~\cite{PID, PyPRC}.

The $d(\gamma,p\vec{n})$ events of interest consist of a primary proton track and a reconstructed neutron, which undergoes a $(n,p)$ charge-exchange reaction in the graphite to produce a secondary proton which gives signals in the MWPC and CB. The primary proton was identified using the correlation between the energy deposits in the PID and CB using $\Delta E-E$ analysis~\cite{PID} along with an associated charged track in the MWPC. The intercept of the primary proton track with the photon beamline allowed determination of the production vertex, and hence permitted the yield originating from the target cell windows to be removed. Neutron $^{12}$C$(n,p)$ charge exchange candidates required an absence of a PID-POL signal on the reconstructed neutron path, while having an associated track in the MWPC and signal in the CB from the scattered secondary proton. The incident neutron angle ($\theta_{n}$) was determined using $E_{\gamma}$ and the production vertex coordinates. A distance of closest approach condition was imposed to ensure  a crossing of the (reconstructed) neutron track and the secondary proton candidate track (measured with MWPC and CB). Once candidate proton and neutron tracks were identified, a kinematic fit was employed to increase the sample purity and improve the determination of the reaction kinematics~\footnote{with photon energy treated as fixed, primary proton - measured, primary neutron unmeasured}, exploiting the fact that the disintegration can be constrained with measurements of two kinematic quantities while three ($\theta_{p}, T_{p} $ and $ \theta_{n}$) are measured in the experiment. A 10\% cut on the probability function was used to select only events from the observed uniform probability region~\cite{mbMainz}.


\section{Determination of neutron polarisation}
The neutron polarisation was determined through analysis of the neutron-spin dependent $^{12}$C$(n,p)$ reactions occuring in the graphite polarimeter. The spin-orbit component of the nucleon-nucleon interaction results in a $\phi$-anisotropy in the produced yield of secondary protons. For a fixed nucleon energy, the secondary proton yield as a function of polar ($\Theta$) and azimuthal ($\phi)$ scattering angle can be expressed as  
\begin{equation}
N(\Theta,\phi)^{pol}=N(\Theta,\phi)^{unpol}\cdot(1+P_n\cdot A_y(\Theta)\cos(\phi)),
\end{equation}
where $P_n$ is the neutron polarisation and $A_y$ is the analysing power. $A_y$ for free $n-p$ scattering is established for the appropriate energy range in the SAID parameterisation~\cite{SAID}. Differences in the analysing power between the free $(n,p)$ process and the in-medium $^{12}$C$(n,p)X$ process were established by a direct measurement of $A_y$ for $^{12}$C$(n,p)X$ by JEDI@Juelich~\cite{Jedi}. Above $T_n=300$~MeV the measured $A_{y}$ agreed with the SAID $(n,p)$ parameterisation to within a few \%. For lower energies, the influence of coherent nuclear processes, such as $^{12}$C$(n,p)^{12}$N resulted in an increased magnitude (around a factor of 2) but exhibiting a similar $\Theta$ dependence to the free reaction~\cite{PyPRC}. The $(n,p)$ analysing power from SAID was corrected \footnote{Note that for the nucleon energies analysed in this experiment the potential contribution of inelastic processes in which an additional pion is produced in the final state are negligible} by the function:

\begin{equation}
A_y(n^{12}C)/A_y(np)=1+e^{(1.82-0.014E_n[MeV])} 
\end{equation}
To reduce systematic dependencies in the simulation of the polarimeter, the events were only retained if $A_y(np)$ was above 0.1, the proton scattering angle ($\Theta$) was in the range $\Theta^{scat}_{p} \in15~-~45^{\circ}$ and $\Theta_n-\Theta^{scat}_{p}> 27^{\circ}$ where $\Theta^{scat}_p$ is the polar angle of scattered proton relative to the direction of the neutron. The latter cut reduced the contribution of secondary protons travelling parallel to the axis of the polarimeter. The scattered yields were corrected for small angle-dependent variations in detection efficiency from the MWPC, established using reconstructed charged particles in the data. The acceptance with the above cuts was determined using a GEANT4~\cite{geant4} simulation of the apparatus. The yield of scattered events was then corrected for this efficiency and the polarisation extracted according to equation 1.   

To quantify systematic errors in the $P^{n}_{y}$ extraction, the analysis cuts were relaxed. This involved widening the cuts on the scattered proton angle and minimum energy (both of which change the MWPC efficiency), reducing the minimum analysing power cut, as well as varying the minimum probability in the kinematic fit up to 40\%~\cite{PyPRC}. The systematic errors are extracted from the resulting variations in the extracted $P^{n}_{y}$, so include significant contributions from the achievable measurement statistics. The main systematic error arose from variations in the $\phi$-dependent detector efficiencies for the secondary protons in GEANT4, which had increasing influence for the lower nucleon energies. The extracted systematic error in $P^{n}_{y}$ was typically around $\pm0.2$ and is presented bin-by-bin with the results in the next section.

\section{Results}

\begin{figure}[!h]
\begin{center}
\includegraphics[width=0.45\textwidth,angle=0]{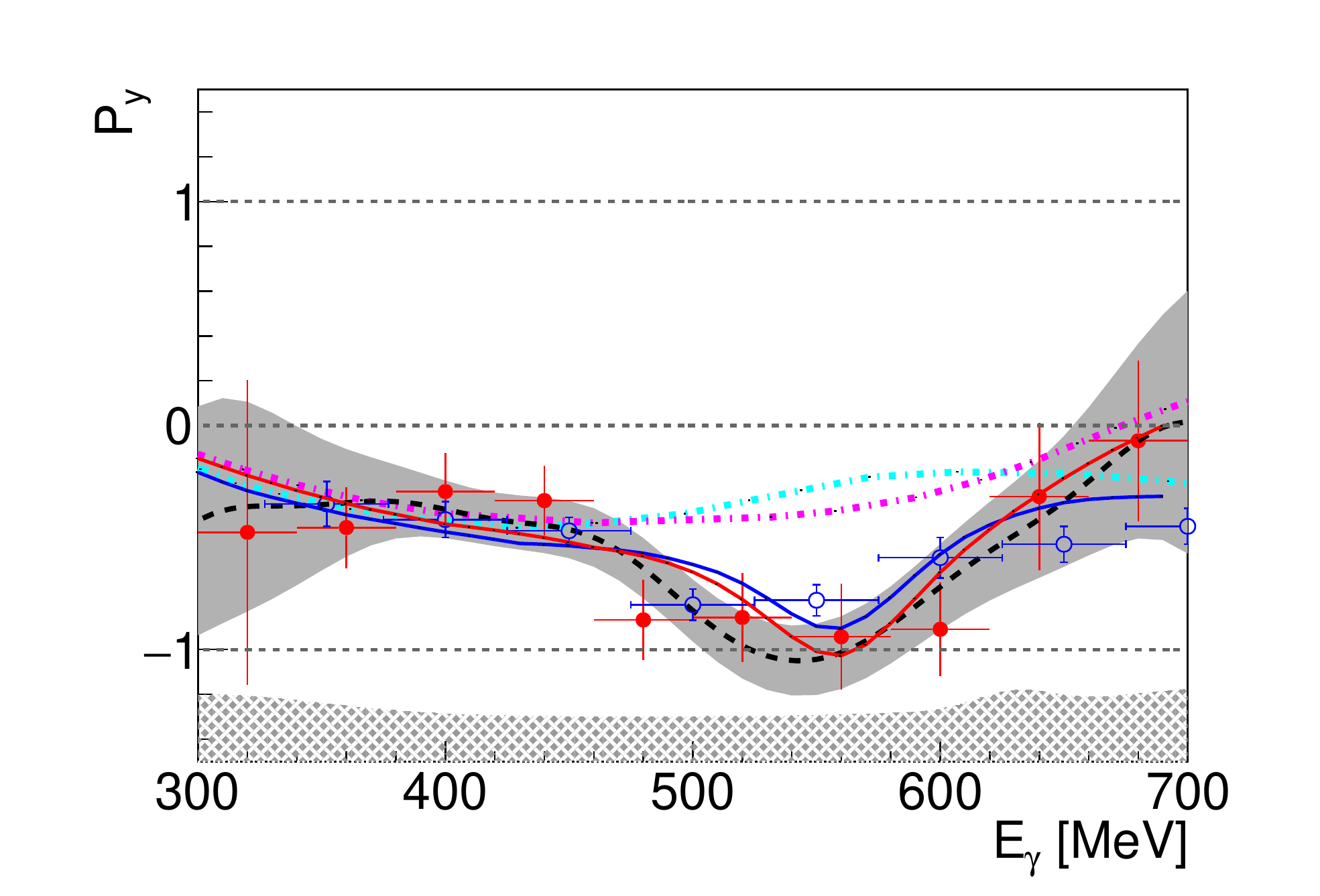}
\end{center}
\caption{$P^{n}_{y}$ (Red filled circles) and previous $P^{p}_{y}$~\cite{TOK1} (blue open circles) for CM angular bins centred on 90$^{\circ}$ as a function of photon energy. The result of an unbinned analysis of $P^{n}_{y}$ is presented as black dashed line with the error bars as a grey band. The dashed-dot lines shows predictions from Ref~\cite{Kang} for $P^{p}_{y}$ (cyan) and $P^{n}_{y}$(pink). The solid lines show the result of the fit with an additional $d^{*}(2380)$ contribution (see text) for $P^{p}_{y}$ (blue) and $P^{n}_{y}$ (red). Systematic uncertainties for the $P^{n}_{y}$ data are shown by the hatched area.}
\label{PyEner}
\end{figure}

The extracted $P_{y}^{n}$ are presented as a function of photon energy at a fixed $\theta_n^{CM}\sim~90^{\circ}$ bin in Fig.~\ref{PyEner}. The $P_{y}^{n}$ observable was extracted in both a binned (red filled circles) and an unbinned (black dashed line) ansatz. Both methods gave consistent results within the statistical accuracy of the data. At the lower photon energies, in the region of the $\Delta$ resonance, $P_{y}^n$ is negative in sign, small in magnitude and rather uniform. However, at higher photon energies the $P_{y}^{n}$ data exhibits a pronounced and sharp structure reaching $\sim~-1$ around $E_{\gamma}\sim$~550~MeV. 
The new data reveals a striking consistency between $P_{y}^n$ and the previous $P_{y}^{p}$~\cite{TOK1} measurements (blue open circles) in the region of the $d^{*}(2380)$.  

The cyan (pink) dashed-dot curves show theoretical calculations of $P_{y}^p$ ($P_{y}^{n}$) respectively. The model includes meson exchange currents ($\pi,\rho,\eta,\omega$) and conventional nucleon resonance degrees of freedom~\cite{Kang}. These calculations reproduce the measured $P_{y}^{p}$ and $P_{y}^{n}$ in the $\Delta$ region, but fail to describe the pronounced and narrow structure centred around $E_{\gamma}\sim$~550~MeV for either observable. At the very highest photon energies the model reproduces the trend indicated in the data, towards smaller magnitudes of $P_{y}^{n}$ compared to $P_{y}^{p}$, an effect  attributed~\cite{Kang} to the $N^*(1520)$ resonance having opposite sign for photocoupling to the neutron and proton~\cite{d13}. Clearly, data at higher $E_{\gamma}$ would help to confirm this trend and better constrain the role of this resonance.

The blue (red) solid lines show a simple approximation to include an additional contribution to these theoretical predictions from the $d^*(2380)$ hexaquark in $P_{y}^{n}$ ($P_{y}^{p}$), taking the established mass and width, and having a magnitude fitted to reproduce the $P_{y}^{p}$ data alone. Previous observations of a lack of mixing of the $d^*(2380)$ with nucleon resonance backgrounds in the inverted reaction $\vec{n}p\to d^*$ gives some justification to this approximate ansatz~\cite{MBE1,MBE2}.  

The main features of the data in the $d^*(2380)$ region, specifically the minima position and width of the dip evident in both $P_{y}^{n}$ and $P_{y}^{p}$, appear consistent with a $d^*(2380)$ contribution of a common magnitude for both channels. Such a common magnitude may be expected from a symmetric decay to $pn$ from a particle which does not mix significantly with other (non-$d^*(2380)$) background contributions. Clearly, more detailed theoretical calculations including the $d^*(2380)$ in a consistent framework within the model would be a valuable next step and we hope our new results will encourage such efforts.  

\begin{figure}[!h]
\begin{center}
\includegraphics[width=0.45\textwidth,angle=0]{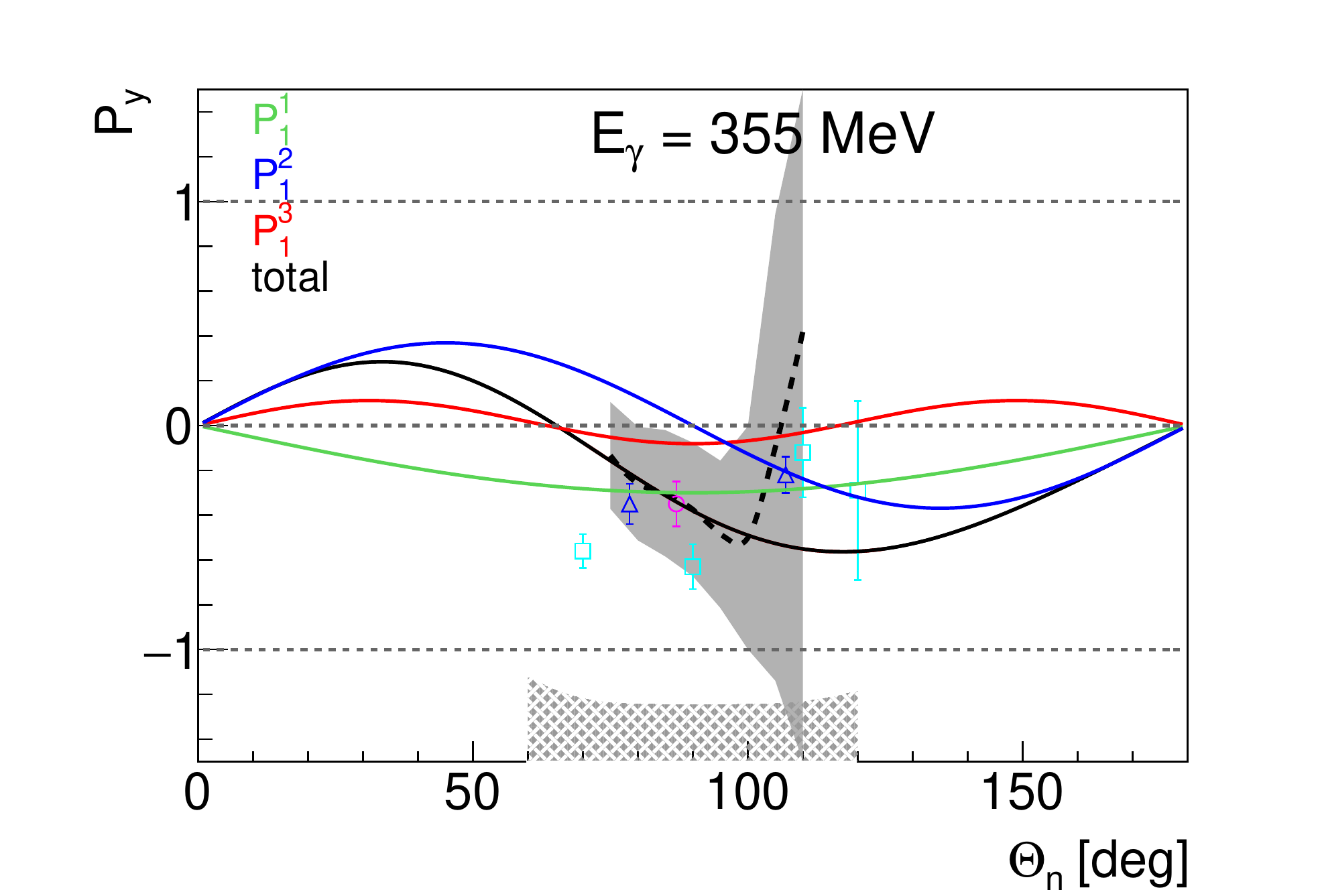}
\includegraphics[width=0.45\textwidth,angle=0]{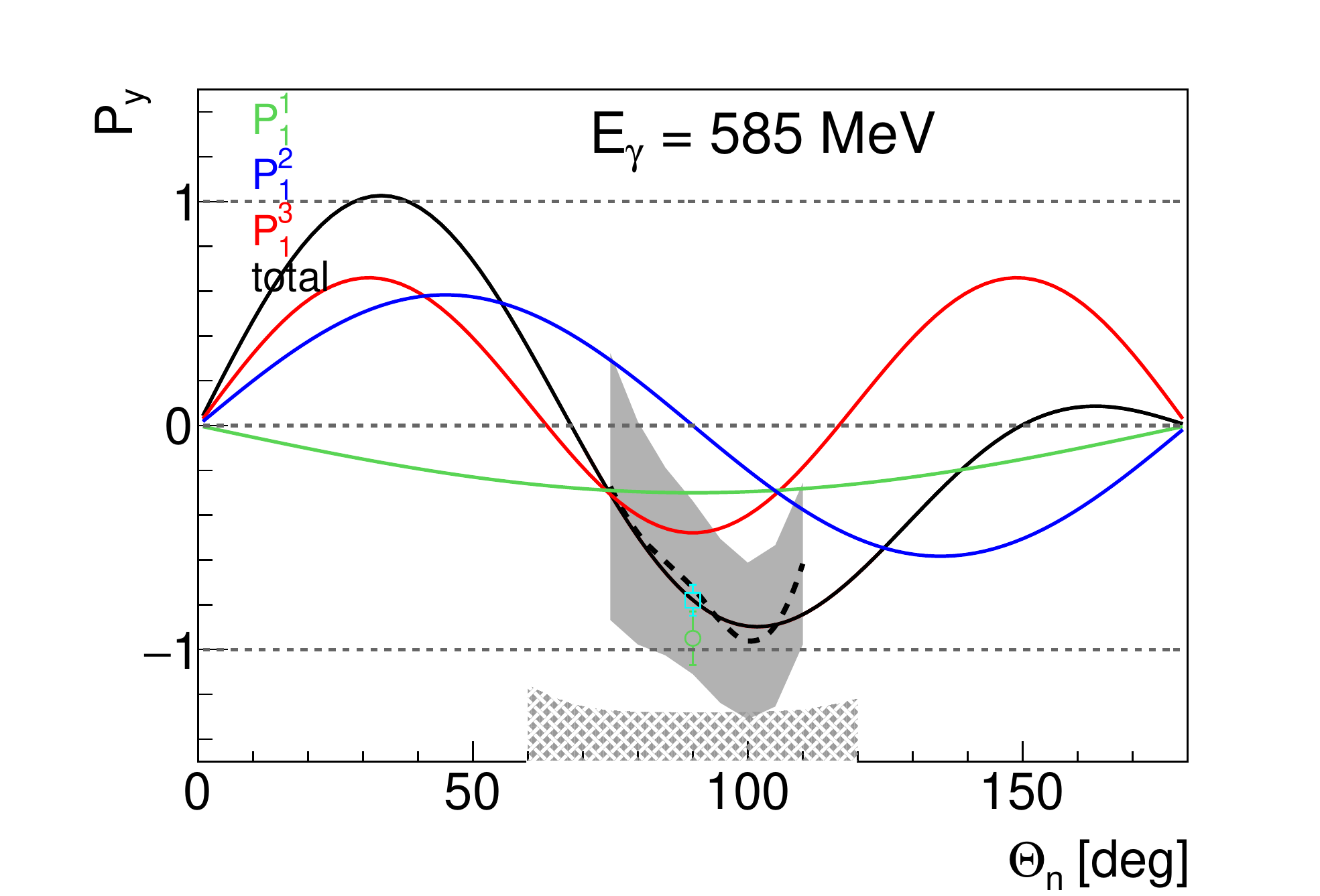}
\end{center}
\caption{$P^{n}_{y}$ is presented as black dashed line with the error bars as a grey band(unbinned ansatz) as a function of $\theta_{n}^{CM}$ for $E_{\gamma}$ bins centred on 355~MeV (upper) and 585~MeV (lower). Existing $P^{p}_{y}$ data are shown as open symbols: cyan squares~\cite{TOK1}, pink circle~\cite{Khar1}, blue triangles~\cite{StanP} and green circle~\cite{JLabP}. The curves are results of the Legendre decomposition (see text): $a_1P^1_1$ (green), $a_2P^2_1$ (blue), $a_3P^3_1$ (red) and their sum (black). Systematic uncertainty is shown as the hatched area.}
\label{PyAng}
\end{figure}

The new data set also has sufficient kinematic acceptance and statistical accuracy to provide a first measurement of the angular dependence of $P_{y}^{n}$.  For a $d^*(2380)\rightarrow pn$ decay, $P_{y}^{n}$ would be expected to exhibit the angular behaviour of the associated $P^3_1$ Legendre function~\footnote{Such analagous polarisation dependencies have already been exploited in $\vec{n}p$-elastic scattering but in inverse direction Ref.~\cite{MBE1,MBE2}, providing key indications of the existence of $d^*(2380)$ in NN scattering.}, reaching a maximum at $\theta^{CM}_{n}=90^{\circ}$ with zero crossings at  $\theta^{CM}_{n}=64^{\circ}$ and $116^{\circ}$. In Fig.~\ref{PyAng}, $P_{y}^{n}$ is presented as a function of $\theta^{CM}_{n}$ for two $E_{\gamma}$ bins, one in the $\Delta$ region and one in the region of the $d^*(2380)$. The $P_{y}^{n}$ from the $\Delta$ region show a broadly flat distribution with $\theta^{CM}_{n}$. In the $d^*(2380)$ region, $P_{y}^{n}$ is higher in magnitude and exhibits a distinct angular dependence with a minima of $\sim~-1$ reached at $\sim$~90$^{\circ}$. There are no previous angular data on $P_{y}^{n}$ in this region. However, the sparse data on $P_{y}^{p}$ (open points) appear consistent with the new $P_{y}^{n}$ measurements in the $\Delta$ region, as predicted by the theoretical model~\cite{Kang}. 


To quantify the dependence of $P^{n}_{y}$ on photon energy and polar angle, we performed an expansion of our results into associated Legendre functions.
\begin{equation}
  P_{y}^{n}=\sum\limits_{l=1}^3 {a_lP_1^l}.
\end{equation}
The result of this expansion can be seen in Fig.~\ref{PyAng}, which shows the fitted contributions from $a_1P_1^1$ (green line), $a_2P_1^2$ (blue line) and $a_3P_1^3$ (red line), and the sum of all contributions (black line). The strongly varying angular behaviour in the $d^*(2380)$ region is consistent with a sizeable $P_1^3$ contribution~\footnote{Note the $P_1^2$ term is exactly zero at $90^{\circ}$ so the energy dependence of Fig.~\ref{PyEner} corresponds to the sum $P_1^1$ and $P_1^3$ only.}.

\begin{figure}[!h]
\begin{center}
\includegraphics[width=0.48\textwidth,angle=0]{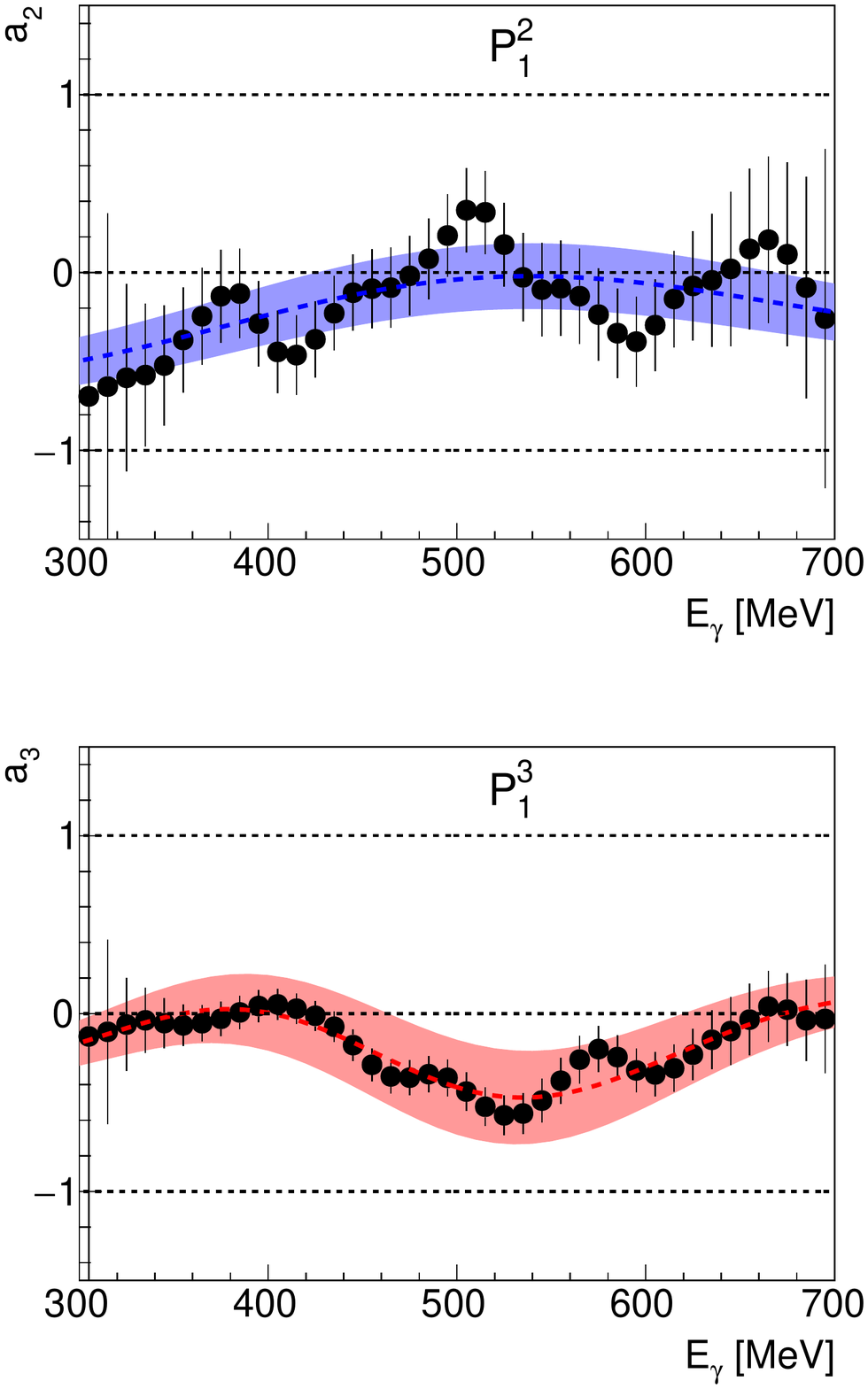}
\end{center}
\caption{Legendre polynomials decomposition of neutron polarisation. $P^2_1$(top) and $P^3_1$(bottom). The $P^1_1$ contribution was fixed to constant value -0.3. The black markers correspond to single energy solutions. The dotted lines represents energy-dependent solutions, with fit errors shown by the coloured bands.}
\label{PyLeg}
\end{figure}

Figure~\ref{PyLeg} shows the $E_{\gamma}$ dependence of fitted expansion coefficients. We employ the prescription adopted in Ref.~\cite{mbMainz} and use two fit methods: (i) a single-energy procedure in which the fit was performed using data from each photon energy bin in isolation (black data points) and (ii) an energy-dependent procedure where the expansion coefficients, $a_l$, were assumed to vary smoothly from bin to bin (dotted lines with the errors represented by bands). The $a_1$ coefficient did not show any particular energy dependent variation, so it was fixed to the value of $a_1=-0.3$. The extracted coefficients are presented as a function of photon energy in Fig.~\ref{PyLeg}. 
The energy dependence of the $P^3_1$ coefficient is consistent with the established mass and width of the $d^*(2380)$ hexaquark ($M=2380\pm 10$~MeV and $\Gamma = 70\pm 10$~MeV), indicating the angular dependence of $P_{y}^{n}$ is consistent with a sizable $J=3$ contribution having properties consistent with those of the $d^*(2380)$.


\section{\label{sec:final} Summary}

The recoil neutron polarisation in deuteron photodisintegration has been measured for $300<E_\gamma<700$~MeV and photon-deuteron centre-of-mass breakup angles for the proton of 60-120$^{\circ}$, providing the first measurement of this fundamental observable at photon energies where the quark substructure of the deuteron can play a role in the mechanism. At lower photon energies, the data are well described by a reaction model which includes meson exchange currents and the known nucleon resonances. At higher photon energies, a narrow structure centred around $E_{\gamma}\sim$~550~MeV is observed in which the neutrons reach a high polarisation. Such behaviour is not reproduced by the theoretical model and is consistent with the ``anomalous" structure observed previously for the recoil proton polarisation~\cite{TOK1}. In a simple ansatz the photon energy and angular dependencies of this ``anomaly" are consistent with a contribution from the $J^{p} = 3^{+}$ $d^*(2380)$ hexaquark.

\section{Acknowledgement}
We are indebted to M. Zurek for providing us data on $n^{12}C$ analysing powers. This work has been supported by the U.K. STFC (ST/L00478X/1, ST/T002077/1, ST/L005824/1, 57071/1, 50727/1 ) grants, the Deutsche Forschungsgemeinschaft (SFB443, SFB/TR16, and SFB1044), DFG-RFBR (Grant No. 09-02-91330), Schweizerischer Nationalfonds (Contracts No. 200020-175807, No. 200020-156983, No. 132799, No. 121781, No. 117601), the U.S. Department of Energy (Offices of Science and Nuclear Physics, Awards No. DE-SC0014323, DEFG02-99-ER41110, No. DE-FG02-88ER40415, No. DEFG02-01-ER41194) and National Science Foundation (Grants NSF OISE-1358175; PHY-1039130, PHY-1714833, No. IIA-1358175), INFN (Italy), and NSERC of Canada (Grant No. FRN-SAPPJ2015-00023).

\end{document}